\documentclass[12pt]{amsart}
\usepackage[dvips]{color}
\usepackage{amsmath}
\usepackage{amsxtra}
\usepackage{amscd}
\usepackage{amsthm}
\usepackage{amsfonts}
\usepackage{amssymb}
\usepackage{eucal}
\usepackage{epsfig}
\usepackage{graphics}

\def\({\left(}
\def\){\right)}

\newcommand{\mub}{\mbox{\boldmath$\mu$}}
\newcommand{\vphis}{\scalebox{.7}{\boldmath$\varphi$}}
\newcommand{\vphi}{\mbox{\boldmath$\varphi$}}
\newcommand{\betab}{\mbox{\boldmath$\beta$}}
\newcommand{\gammab}{\mbox{\boldmath$\gamma$}}

\newcommand{\nn}{\nonumber}
\newcommand{\bea}{\begin{eqnarray}}
\newcommand{\ena}{\end{eqnarray}}
\def\bel{\begin{eqnarray}}
\def\enl{\end{eqnarray}}
\newcommand{\be}{\begin{eqnarray*}}
\newcommand{\en}{\end{eqnarray*}}
\newcommand{\ba}{\begin{array}}
\newcommand{\ea}{\end{array}}

\newcommand{\C}{{\mathbb C}}
\newcommand{\Z}{{\mathbb Z}}

\newenvironment{tenumerate}{
  \begin{enumerate}
  
  }{\end{enumerate}}
\newcommand{\bi}{\begin{tenumerate}}
\newcommand{\ei}{\end{tenumerate}}
\newcommand{\isoto}[1][]%
{{\mathop{\buildrel{\sim}\over\longrightarrow}\limits_{#1}}}

\def\[{\left[}
\def\]{\right]}
\newcommand{\la}{\lambda}

\newcommand{\al}{\alpha}

\numberwithin{equation}{section}

\newcommand{\gab}{\mbox{\boldmath$\gamma $}}
\newcommand{\beb}{\mbox{\boldmath$\beta $}}

\def\bi{\mathbf{i}}

\newcommand{\zbz}{z,\bar{z}}

\begin{document}

\hfill TCDMATH 11-03

\hfill HMI 11-01
\vskip 1cm

\begin{title}{
Fermionic 
screening operators \\ in the sine-Gordon model}
\end{title}
\author{M.~Jimbo, T.~Miwa and  F.~Smirnov}
\address{MJ: Department of Mathematics, 
Rikkyo University, Toshima-ku, Tokyo 171-8501, Japan}
\email{jimbomm@rikkyo.ac.jp}
\address{TM: Department of 
Mathematics, Graduate School of Science,
Kyoto University, Kyoto 606-8502, 
Japan}\email{tmiwa@math.kyoto-u.ac.jp}
\address{
FS:\footnote
{Membre du CNRS}
Hamilton Mathematical Institute and School of Mathematics,
Trinity College, Dublin 2, Ireland} \address{
Laboratoire de Physique Th{\'e}orique et
Hautes Energies, Universit{\'e} Pierre et Marie Curie,
Tour 13, 4$^{\rm er}$ {\'e}tage, 4 Place Jussieu
75252 Paris Cedex 05, France}\email{smirnov@lpthe.jussieu.fr}

\begin{abstract}
Extending our previous construction in the sine-Gordon model,  
we show how to introduce 
two kinds of fermionic 
screening operators, in close analogy with conformal field theory with $c<1$.
\end{abstract}

\maketitle

\definecolor{purple}{rgb}{0,0,0}
\definecolor{oldred}{rgb}{0,0,0}
\definecolor{brown}{rgb}{0.8,0.3,0.3}
\definecolor{2/20}{rgb}{0,0,0}
\definecolor{2/21}{rgb}{0,0,0}
\definecolor{2/23}{rgb}{0,0,0}
\definecolor{3/2}{rgb}{0.8,0,0.2}

\section{Introduction}

In our previous work \cite{HGSV}, we have introduced and studied 
certain fermions which act on the space of local fields in the sine-Gordon model. 
The present note is intended as a supplement to that paper. 
Our aim here is to explain how to introduce fermionic analogs of the two screening operators 
well-known in CFT with $c<1$. 

In our notation, the Euclidean action of the sine-Gordon model is 
\begin{align}
\mathcal{A}^\mathrm{sG}
&=\int \left\{ \Bigl[\frac 1 {4  \pi} 
\partial _z\vphi (z,\bar{z})\partial_{\bar{z}}\vphi (z,\bar{z})-
\frac{\mub ^2}{\sin\pi\beta ^2} e^{-i\beta\vphis(z,\bar{z})}\Bigr]\right.
\label{action1}
\\ 
&\quad\quad\quad\quad\quad\quad\quad\quad\quad\quad\quad\quad\quad \left.
-
\frac{\mub ^2}{\sin\pi\beta ^2}   
e^{i\beta\vphis(z,\bar{z})}    
\right\}\frac{idz\wedge d\bar{z}}2\,.
\nn
\end{align}
In place of the coupling parameter $\beta^2$, we shall
henceforth  use the parameter
\begin{align}
\nu =1-\beta ^2\,,
\label{nu}
\end{align}
assuming $1/2<\nu<1$. 
We view the sine-Gordon model as a perturbation of the Liouville CFT.
The latter is given by the action \eqref{action1} without the last term, 
and the central charge is
\begin{align}
c=1-\frac{6\nu^2}{1-\nu}\,.
\label{central}
\end{align}
The exponential field parametrised as
\begin{align}
\Phi_\al(\zbz)=e^{
\ \frac { \nu}{2(1-\nu)}
\al \left\{ {i\beta} \vphis (\zbz)\right\}}\, 
\label{primary}
\end{align}
turns in the CFT limit into a primary field 
with the scaling dimension
\begin{align}
\Delta_\al=\frac{\nu^2}{4(1-\nu)}\al(\al-2)\,.
\label{Delta}
\end{align}  
Abusing the language, we refer to \eqref{primary} 
as a primary field in the sine-Gordon model as well.  

In \cite{HGSV}, we have considered two types of fermions.
One of them, denoted $\betab^*_{2j-1},\gammab^*_{2j-1}$ and 
$\bar{\betab}^*_{2j-1},\bar{\gammab}^*_{2j-1}$, are local in nature. 
Acting on a primary field \eqref{primary}, 
together with the action of local integrals of motion, 
they create the space of all descendant fields, which in the CFT limit  
corresponds to the Verma module. 
The other type, denoted 
$\betab^*_{\mathrm{screen},j},\gammab^*_{\mathrm{screen},j}$ 
and  $\bar{\betab}^*_{\mathrm{screen},j},\bar{\gammab}^*_{\mathrm{screen},j}$,  
are non-local. We call them the first fermionic screening operators. 
They relate primary fields with shifted exponents,
\begin{align*}
\Phi_{\al+2n\frac{1-\nu}{\nu}}
\quad (n\in \Z_{\ge0}),
\end{align*} 
as well as their corresponding descendants. 
For a more precise statement, see section \ref{sec:CFTdes}
below. 
These are analogous to one of the two screening operators which exist 
in Liouville CFT. 
For these primary fields and fermionic descendants, the vacuum expectation values (VEVs)
can be written explicitly. 
This is the main advantage of the fermionic description of the space of local fields.

As is well known, in CFT there exist another set of screening operators 
due to the symmetry
\begin{align*}
1-\nu~\longleftrightarrow~ \frac{1}{1-\nu}\,.
\end{align*}
The goal of this paper is to introduce their fermionic
counterpart. 
We shall call them the second fermionic screeing operators and denote them by
$\betab^*_{\mathrm{SCREEN},j},\gammab^*_{\mathrm{SCREEN},j}$ 
and  $\bar{\betab}^*_{\mathrm{SCREEN},j},\bar{\gammab}^*_{\mathrm{SCREEN},j}$.   

The text is organised as follows. 
In Section 2, we consider the case of CFT on a cylinder. 
In \cite{HGSV}, fermions are introduced by specifying their three point functions, 
which are encoded in a `structure function' called $\omega^\mathrm{sc}$. 
In section 2.1 we re-examine the analytic structure of $\omega^\mathrm{sc}$ and 
find its asymptotic behavior in sectorial domains. 
In section 2.2 we define the operators 
$\betab^*_{\mathrm{SCREEN},j}$, etc., from the asymptotic expansion of 
$\omega^\mathrm{sc}$. 
They relate the primary fields 
\begin{align*}
\Phi_{\al-2m\frac{1}{\nu}}
\quad (m\in\Z_{\ge0}).
\end{align*} 
We formulate the conjectural formulas pertaining to primary fields, 
and in section 2.3 for their descendants. 
In section 2.4 we verify various consistency relations 
required from the conjectures.
In section 2.5 we glue togehter the two chiralities and
check their three point functions against the
Zamolodchikov-Zamolodchikov formula. 

Section 3 is devoted to the case of the sine-Gordon model. 
Following the method in CFT we introduce the second fermionic screening operators
by giving formulas for the VEV. 
In section 3.2 we check the formula with the Lukyanov-Zamolodchikov formula for 
one-point functions in the sine-Gordon model on the infinite plane. 
In section 3.3 we verify the validity of the consistency relations for them. 


%

\section{Fermionic screening operators in CFT}

In this section we explain how to introduce the second fermionic 
screening operators in CFT. 

\subsection{Analytic structure of $\omega^{\mathrm{sc}}$}

In \cite{HGSIV} we considered chiral CFT on a cylinder $\C/2\pi i \Z$, 
with the central charge \eqref{central},  
and primary fields $\phi_\al(x)$ of scaling dimension \eqref{Delta}.
The objects of interest are the fermionic operators acting on 
the space of descendants of $\phi_\al(0)$.  
They are defined through the ``structure function'' $\omega^\mathrm{sc}(\la,\mu|\kappa,\kappa',\al)$.  
It describes the three point functions of the fermioninc descendants of $\phi_\al(0)$,   
in the presence of primary fields 
$\phi_{1-\kappa'}(-\infty)$, $\phi_{1+\kappa}(\infty)$ 
inserted at $x=\pm\infty$. 
In this subsection, we re-examine the analytic structure of $\omega^\mathrm{sc}$ in order to 
extract the fermionic screening operators. 
So far the asymptotic analysis has been carried out only in the case $\kappa'=\kappa$.    
Throught this text we shall make this assumption.

It is convenient to slightly modify $\omega^\mathrm{sc}$ in \cite{HGSIV}, (11.5), 
and start from the function
\begin{align}
&\omega^\mathrm{sc}_0(\la,\mu|\kappa,\kappa,\al)=\frac{1}{2\pi i}\int\!\!\int dl\,dm\,
\tilde{S}(l,\al)\tilde{S}(m,2-\al)\label{omega-sc}\\
&\quad\times\Theta(l-i0,m|\kappa,\al)\,
\bigl(e^{\pi i \nu+\delta}t\bigr)^{il}\bigl(e^{\pi i \nu+\delta}u\bigr)^{im}\,.\nn
\end{align}
Here we have set
\begin{align*}
&\la^2=c(\nu,\kappa)t,\ \mu^2=c(\nu,\kappa)u,
\\
&c(\nu,\kappa)=\Gamma(\nu)^{-2}e^{\delta}\Bigl(\frac{\nu\kappa}{2}\Bigr)^{2\nu},\quad
\delta=-\nu\log\nu-(1-\nu)\log(1-\nu),
\end{align*}
and
\begin{align*}
&\tilde{S}(k,\al)=
\frac{\Gamma\bigl(-ik+\frac{\al}{2}\bigr)
\Gamma\bigl(\frac{1}{2}+i\nu k\bigr)}
{\Gamma\bigl(-i(1-\nu)k+\frac{\al}{2}\bigr)
\sqrt{2\pi}(1-\nu)^{(1-\al)/2}}
\,.
\end{align*}
The function $\Theta(l,m|\kappa,\al)$ is essentially
the Mellin transform of the resolvent kernel of a linear integral operator 
associated with the Thermodynamic Bethe Ansatz equation. 
It has an asymptotic expansion in $\kappa^{-2}$  
\begin{align*}
\Theta(l,m|\kappa,\al)\simeq \sum_{n=0}^\infty \Theta_n(l,m|\al)\kappa^{-2n}\,,
\quad \Theta_0(l,m|\al)=\frac{-i}{l+m}\,.
\end{align*}
The coefficients $\Theta_n(l,m|\al)$ ($n\ge1$) are 
polynomials in $l,m$  determined from the recursion relation in \cite{HGSIV}, Section 11. 
We remark that, 
when the fields $\phi_{1\mp\kappa}(\pm\infty)$
are changed to descendants,   
the same structure for $\omega^{\mathrm{sc}}_0$ still persists,  
the only change 
being in $\Theta(l,m|\kappa,\al)$ (see e.g. \cite{Boos}).

Since 
\begin{align*}
\bigl(e^{\pi i\nu+\delta}\bigr)^{ik}\tilde{S}(k,\al)
\simeq
\begin{cases}
e^{-2\pi\nu k}& (k\to \infty),\\
1 & (k\to -\infty),\\
\end{cases}
\end{align*}
the integral \eqref{omega-sc} converges term by term in $\kappa^{-2}$,  
provided $t,u$ belong to the domain 
\begin{align}
&\mathcal{S}_+~:~ -2\pi \nu<\arg t<0\,.\label{CONVDOM}
\end{align}
By moving the contours into the lower half plane and taking residues in $l,m$, 
one obtains an asymptotic expansion of the form
\begin{align}
\omega^\mathrm{sc}_0(\la,\mu|\kappa,\kappa,\al)
\simeq \sum_{j,k=1}^\infty t^{\al/2+j-1}u^{-\al/2+k}\omega^\mathrm{sc}_{0,j,k}(\kappa,\al)\,
\quad (t,u\to 0,\ t,u\in \mathcal{S}_+).
\label{o0Ser}
\end{align}
It is easy to see that 
the series in the right hand side is actually convergent on  $D\times D$, where 
$D=\{t\in\C\mid |t|<1\}$ (see Fig.1).

\vskip .5cm
\begin{center}
\includegraphics[scale=0.5]{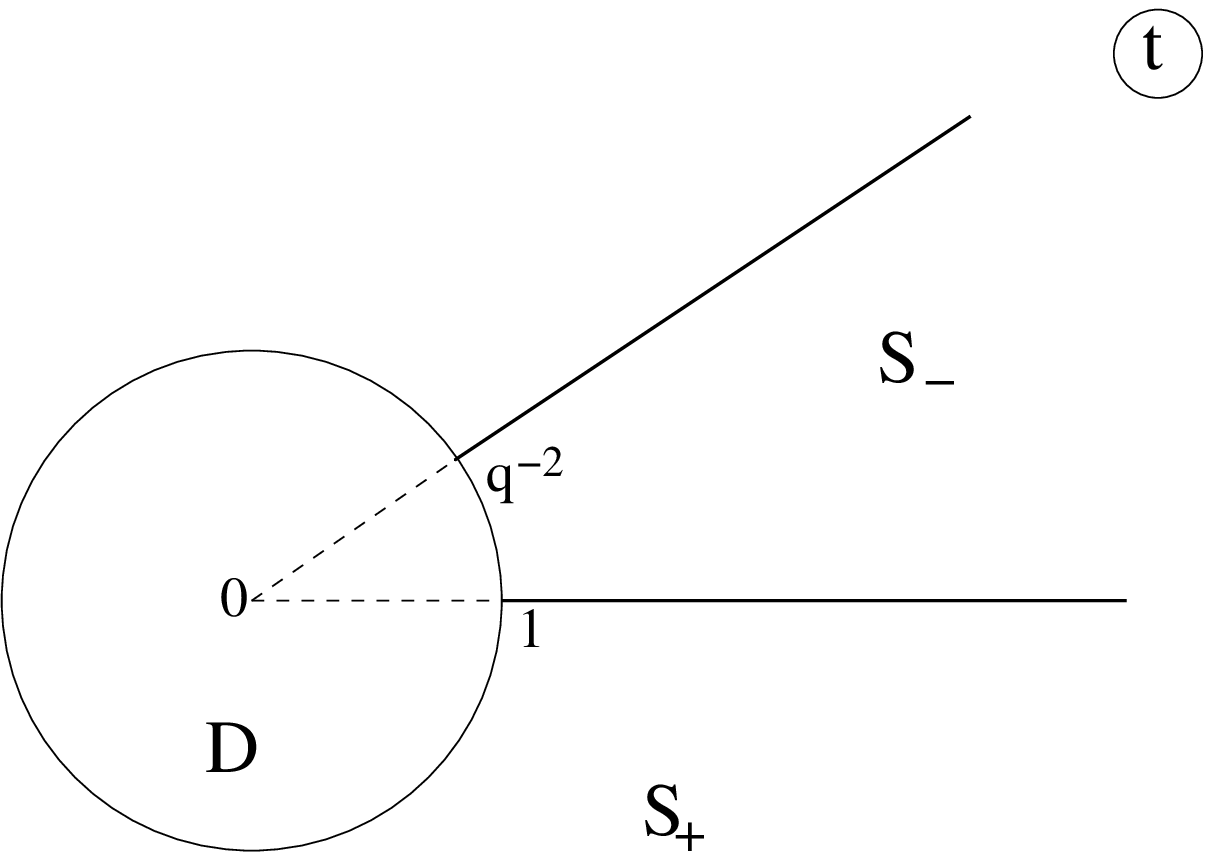}
\end{center}

{\it Fig.1: Domains of analyticity of $\omega^{\mathrm{sc}}_0$. 
This function is holomorphic in the unit disk $D$, and is continued 
analytically to the complex plane 
with the two cuts $[1,\infty)$ and $[q^{-2},q^{-2}\infty)$ ($q=e^{\pi i \nu}$) being
removed. 
}
\vskip .5cm

In fact,  $\omega^{\mathrm{sc}}_0$ 
can be continued analytically in each variable 
to the entire complex plane with the two cuts
$[1,\infty)$, $[q^{-2},q^{-2}\infty)$  ($q=e^{\pi i \nu}$) being removed. 
The analytic continuation in the variable $t$ is achieved explicitly
by a similar integral \eqref{omega-sc}, wherein 
the function $\tilde S(l,\al)$ is changed to $\tilde S(l,\al)A_-(l,\al)$ with 
\begin{align*}
A_-(l,\al)=-\frac{\sinh(\pi \nu l)}{\sinh\pi\bigl((1-\nu)l+i\frac{\al}{2}\bigr)}
e^{\pi(l+\frac {i\al} 2)}\,.
\end{align*}
To see this, notice that  
the series expansion \eqref{o0Ser} remains intact under this modification 
because  
\begin{align*}
&A_-\bigl(i(j-\frac{\al}{2}),\al\bigr)=1\quad (j\in \Z)\,.
\end{align*}
On the other hand, due to the behavior
\begin{align*}
&A_-(l,\al)\simeq
\begin{cases}
e^{2\pi\nu l} & (l\to \infty)\\
e^{2\pi (1-\nu)l}& (l \to -\infty)\\
\end{cases}\,
\end{align*}
the domain of convergence of the integral is changed from $\mathcal{S}_+$ to 
\begin{align*}
\mathcal{S}_-~:~ 0<\arg t< 2\pi (1-\nu)\,.
\end{align*}

Of course the same procedure applies to the  $u$ variable as well. 
Altogether one obatins 4 different expressions of the same function
$\omega_0^{\mathrm{sc}}$ corresponding to 
the sectors $\mathcal{S}_\epsilon\times \mathcal{S}_{\epsilon'}$, $\epsilon,\epsilon'=\pm$. 
Let us write 
\begin{align*}
\omega_0^{\mathrm{sc}}(\la,\mu|\kappa,\kappa,\al)
&=
\omega_{\epsilon,\epsilon'}^{\mathrm{sc}}(\la,\mu|\kappa,\kappa,\al)
+
\omega_{0,\epsilon,\epsilon'}(\la/\mu|\al)\,,
\end{align*}
where 
\begin{align*}
&\omega_{\epsilon,\epsilon'}^{\mathrm{sc}}(\la,\mu|\kappa,\kappa,\al)
\\
&=\frac{1}{2\pi i}\int\!\!\int dl dm
\tilde{S}_\epsilon(l,\al)\tilde{S}_{\epsilon'}(m,2-\al)
\Theta(l+i0,m|\kappa,\al)\,
\bigl(e^{\pi i \nu+\delta}t\bigr)^{il}\bigl(e^{\pi i \nu+\delta}u\bigr)^{im}\,,\\
&\tilde{S}_\pm(k,\al)=\tilde{S}(k,\al)A_\pm(k,\al)\,,
\quad
A_+(k,\al)=1\,,
\end{align*}
and
\begin{align*}
\omega_{0,\epsilon,\epsilon'}(\zeta|\al)
&=-i\int dl\, \zeta^{2il}\tilde{S}_\epsilon(l,\al)\tilde{S}_{\epsilon'}(-l,2-\al)\,.
\end{align*}
The function $\omega^{\mathrm{sc}}_{+,+}$ coincides with $\omega^{\mathrm{sc}}$ 
considered originally in \cite{HGSIV}. 

When $\la^2,\mu^2\to\infty$,
the modified function $\omega^\mathrm{sc}_{\epsilon,\epsilon'}$ has 
an asymptotic  expansion in each sector.  
For instance, one has the expansions of the form 
\begin{align}
\omega_{+,+}^{\mathrm{sc}}(\la,\mu|\kappa,\kappa,\al)
&\simeq 
\sum_{j,k=1}^\infty \la^{-\frac{2j-1}{\nu}}\mu^{-\frac{2k-1}{\nu}}
\omega_{+,+;j,k}^{\mathrm{sc}}(\kappa,\al)\,,
\label{om++}\\
\omega_{-,+}^{\mathrm{sc}}(\la,\mu|\kappa,\kappa,\al)
&\simeq 
\sum_{j,k=1}^\infty \la^{-\frac{2j-1}{\nu}}\mu^{-\frac{2k-1}{\nu}}
\frac{e^{\frac{\pi i}{2\nu}(2j-1+\nu\al)}}
{\cos\frac{\pi i}{2\nu}(2j-1+\nu\al)}
\omega_{+,+;j,k}^{\mathrm{sc}}(\kappa,\al)\,
\label{om+-}\\
&+\sum_{j,k=1}^\infty \la^{-\frac{2j-\al}{1-\nu}}\mu^{-\frac{2k-1}{\nu}}
\omega_{-,+;j,k}^{\mathrm{sc}}(\kappa,\al)\,,
\nn
\end{align}
in $\mathcal{S}_+\times\mathcal{S}_+$ and 
$\mathcal{S}_-\times\mathcal{S}_+$, respectively.
The same quantity $\omega_{+,+;j,k}^{\mathrm{sc}}(\kappa,\al)$
appears in two places.

Note that $\omega_{0,\epsilon,\epsilon'}(\zeta|\al)$ 
is independent of $\kappa$.
Later on, in Section 3.1, we shall use their asymptotic expansions as $\zeta^2\to\infty$:
\begin{align*}
\omega_{0,++}(\zeta|\al)&\simeq
\sum_{j=1}^\infty 
\zeta^{-\frac{2j-1}{\nu}}t_{2j-1}(\al)
+\sum_{j=1}^\infty \zeta^{-2j+\al}t'_{j}(\al)\,,\\
\omega_{0,\mp\pm}(\zeta|\al)&\simeq
\sum_{j=1}^\infty \zeta^{-\frac{2j-1}{\nu}}
\frac{e^{\frac{\pi i}{2\nu}(2j-1\pm\nu\al)}t_{2j-1}(\al)}
{\cos\frac{\pi}{2\nu}\bigl(2j-1\pm\nu\al\bigr)}
+\sum_{j=1}^\infty \zeta^{-2j+\al}t'_j(\al)\,,\\
\omega_{0,--}(\zeta|\al)&\simeq\sum_{j=1}^\infty \zeta^{-\frac{2j-1}{\nu}}
\frac{e^{\frac{\pi i}\nu(2j-1)}t_{2j-1}(\al)}{\prod_\pm\cos\frac{\pi}{2\nu}(2j-1\pm\nu\al)}\,
+\sum_{j=1}^\infty \zeta^{-2j+\al}t'_j(\al)
+\sum_{j=1}^\infty \zeta^{-\frac{2j-\al}{1-\nu}}t''_j(\al)\,,
\end{align*}
where
\begin{align}
&t_a(\al)=\frac{i}{\nu}\cot\frac{\pi}{2\nu}(a+\nu\al)\,,
\label{tal}\\
&t'_j(\al)=-i\tan\frac{\pi\nu}{2}(2j-\al)\,,
\label{tal2}
\\
&t''_j(\al)=\frac{i}{1-\nu}\tan\frac{\pi\nu(2j-\al)}{2(1-\nu)}\,.
\label{tal1}
\end{align}
The asymptotic expansions as 
$\zeta^2\to0$ are obtained from the equality
\begin{align*}
\omega_{0,\epsilon,\epsilon'}(\zeta|\al)=\omega_{0,\epsilon',\epsilon}(\zeta^{-1}|2-\al).
\end{align*}

\subsection{Second screening operators}
The function $\tilde{S}(k,\al)$ has three series of poles and zeroes in $k$,  
\begin{align}
&\text{poles}\quad \frac{i(2j-1)}{2\nu}\,,\frac{i(2-2j-\al)}{2}\,;
\quad \text{zeros}\quad \frac{i(2-2j-\al)}{2(1-\nu)}\,,
\label{poles}
\end{align}
where $j\in \Z_{\ge1}$. 
In \cite{HGSIV}, the first is used to introduce the fermionic creation operators
\begin{align*}
\betab^*(\la)=\sum_{j=1}^\infty \lambda^{-\frac{2j-1}{\nu}}\betab^*_{2j-1},
\quad
\gammab^*(\la)=\sum_{j=1}^\infty \lambda^{-\frac{2j-1}{\nu}}\gammab^*_{2j-1}, 
\end{align*}
while the second is used for the screening operators 
\begin{align*}
\betab_{\mathrm{screen}}^*(\la)=\sum_{j=1}^\infty \lambda^{2j-2+\al}\betab^*_{\mathrm{screen},j}, 
\quad
\gammab_{\mathrm{screen}}^*(\la)=\sum_{j=1}^\infty \lambda^{2j-\al}
\gammab^*_{\mathrm{screen},j}. 
\end{align*}
They are related to the three point functions as follows.
\begin{align*}
&\langle \betab^*(\la)\gammab^*(\mu)\rangle_{\kappa,\al}
=\omega^{\mathrm{sc}}(\la,\mu|\kappa,\kappa,\al)\quad(\la,\mu\rightarrow\infty)\,,
\\
&\langle \betab^*(\la)\gammab^*_{\mathrm{screen}}(\mu)\rangle_{\kappa,\al}
=\omega^{\mathrm{sc}}(\la,\mu|\kappa,\kappa,\al)\quad(\la\rightarrow\infty,\mu\rightarrow0)\,.
\end{align*}
Here and after we use the abbreviation
\begin{align}
\langle 
\betab^*(\la)\gammab^*(\mu)
\rangle_{\kappa,\al}=
\frac{\langle 1-\kappa|
\betab^*(\la)\gammab^*(\mu)
\phi_\al(0)|1+\kappa\rangle}
{\langle 1-\kappa|\phi_\al(0)|1+\kappa\rangle}\,,
\label{exp}
\end{align}
and so on. We shall refer to them as `pairings' of fermions.

The analysis in the previous subsection motivates us to introduce yet another fermions 
\begin{align*}
&\betab^*_-(\la)=
\tilde\betab^*(\la)
+\betab^*_{\mathrm{SCREEN}}(\la)\,,
\ \gammab^*_-(\la)=
\tilde\gammab^*(\la)
+\gammab^*_{\mathrm{SCREEN}}(\la)\,,\\
&\tilde\betab^*(\la)=
\sum_{j=1}^\infty \la^{-\frac{2j-1}\nu}
\frac{e^{\frac{\pi i}{2\nu}(2j-1+\nu\al)}}{\cos\frac{\pi}{2\nu}(2j-1+\nu\al)}
\cdot \betab^*_{2j-1},\\
&\tilde\gammab^*(\la)=
\sum_{j=1}^\infty \la^{-\frac{2j-1}\nu}
\frac{e^{\frac{\pi i}{2\nu}(2j-1-\nu\al)}}
{\cos\frac{\pi}{2\nu}(2j-1-\nu\al)}
\cdot \gammab^*_{2j-1},
\end{align*}
such that together with $\betab^*_+(\la)=\betab^*(\la)$ and
$\gammab^*_+(\la)=\gammab^*(\la)$ they satisfy
\begin{align*}
\langle \betab^*_\epsilon(\la) \gammab^*_{\epsilon'}(\mu) \rangle_{\kappa,\al}
=\omega^{\mathrm{sc}}_{\epsilon,\epsilon'}(\la,\mu|\kappa,\kappa,\al)\,
\quad (\epsilon,\epsilon'=\pm),
\end{align*}
for $\la,\mu\rightarrow\infty$.
The operators
\begin{align*}
&\betab^*_{\mathrm{SCREEN}}(\la)=
\sum_{j=1}^\infty \la^{-\frac{2j-\al}{1-\nu}}\betab^*_{\mathrm{SCREEN},j}\,,
\quad
\gammab^*_{\mathrm{SCREEN}}(\la)=
\sum_{j=1}^\infty \la^{-\frac{2j-2+\al}{1-\nu}}\gammab^*_{\mathrm{SCREEN},j}
\end{align*}
correspond to the poles of $A_-(k,\al)$ which are not canceled by the zeros in \eqref{poles}.

We assign the following degrees to the Fourier components:
\begin{alignat*}{2}
&\deg \betab^*_{2j-1}=2j-1\,, & 
&\ \deg \gammab^*_{2j-1}=2j-1\,,
\\
&\deg \betab^*_{\mathrm{screen},j}=-\nu(2j-2+\al)\,, & 
&\ \deg \gammab^*_{\mathrm{screen},j}=-\nu(2j-\al)\,,
\\
&\deg \betab^*_{\mathrm{SCREEN},j}=\frac{\nu}{1-\nu}(2j-\al)\,,&
&\ \deg \gammab^*_{\mathrm{SCREEN},j}=\frac{\nu}{1-\nu}(2j-2+\al)\,. 
\end{alignat*}
Their pairings are given explicitly as follows. Set 
\begin{align*}
&\tilde{D}_{2j-1}(\al)=-\frac{\sqrt{i}}{\nu}\frac{1}{(j-1)!}
\frac{\Gamma\bigl(\frac{1}{\nu}(j-\frac{1}{2})+\frac{\al}{2}\bigr)}
{\Gamma\bigl(\frac{1-\nu}{\nu}(j-\frac{1}{2})+\frac{\al}{2}\bigr)},\\
&\tilde{E}_{j}(\al)=-\frac{1}{\sqrt{i}}\frac{(-1)^j}{(j-1)!}
\frac{\Gamma\bigl(\frac{1}{2}+\nu(j-\frac{\al}{2})\bigr)}
{\Gamma\bigl(1-(1-\nu)j-\frac{\nu \al}{2}\bigr)}\cdot e^{\pi i \nu(j-\frac{\al}{2})}\,,\\
&\tilde{F}_{j}(\al)=\frac{1}{\sqrt{i}}\frac{1}{1-\nu}\frac{(-1)^j}{(j-1)!}
\frac{\Gamma\bigl(\frac{1}{2}-\frac{\nu}{1-\nu}(j-\frac{\al}{2})\bigr)}
{\Gamma\bigl(1-\frac{1}{1-\nu}j+\frac{\nu \al}{2(1-\nu)}\bigr)}.
\end{align*}
Then we have 
\begin{align}
&\langle \betab^*_{a}\gammab^*_{b}\rangle_{\kappa,\al}
=\tilde{D}_{a}(\al)\tilde{D}_{b}(2-\al)
\tilde \Theta({\textstyle\frac{ia}{2\nu},\frac{ib}{2\nu}|\kappa,\al})\,,\label{FPAIRING}\\
&\langle \betab^*_{a}\gammab^*_{\mathrm{screen},k}\rangle_{\kappa,\al}
=\tilde{D}_{a}(\al)\tilde{E}_{k}(\al)
\tilde \Theta({\textstyle\frac{ia}{2\nu},-i\left(k-\frac\al2\right)|\kappa,\al})\,,\nn\\
&\langle \betab^*_{\mathrm{SCREEN},j}\gammab^*_{b}\rangle_{\kappa,\al}
=\tilde{F}_{j}(\al)\tilde{D}_{b}(2-\al)
\tilde \Theta({\textstyle\frac i{1-\nu}\left(j-\frac\al2\right),\frac{ib}{2\nu}| \kappa,\al})\,,\nn\\
&\langle \betab^*_{\mathrm{SCREEN},j}\gammab^*_{\mathrm{screen},k}\rangle_{\kappa,\al}
=\tilde{F}_{j}(\al)\tilde{E}_{k}(\al)
\tilde\Theta({\textstyle\frac i{1-\nu}\left(j-\frac\al2\right),-i\left(k-\frac\al2\right)|\kappa,\al})\,.\nn
\end{align}
where
\begin{align*}
&\tilde\Theta(l,m|\kappa,\al)=\Theta(l,m|\kappa,\al)
x(\kappa)^{i(l+m)},\quad
x(\kappa)=\Bigl(\frac{\nu\kappa}{2}\Bigr)^{-2\nu}\Gamma(\nu)^2\,.
\end{align*}
We have given only those pairings which will be relevant to our calculations. 

We remark that the fermionic operators for the anti-chiral CFT 
are introduced by working with the function
\begin{align}
&\bar{\omega}^\mathrm{sc}_0(\la,\mu|\kappa,\kappa,\al)
=\frac{1}{2\pi i}\int\!\!\int dl\,dm
\tilde{S}(l,2-\al)\tilde{S}(m,\al)\label{omega-bar}\\
&\quad\times\Theta(l-i0,m|-\kappa,2-\al)\,
\bigl(e^{-\pi i \nu+\delta}t\bigr)^{il}\bigl(e^{-\pi i \nu+\delta}u\bigr)^{im}\,,\nn
\end{align}
where $c(\nu,\kappa)t=\la^{-2},c(\nu,\kappa)u=\mu^{-2}$,
as a counterpart for $\omega_0^\mathrm{sc}$.
We have the following
(conjectural) symmetry for the coefficients in the asymptotic expansions:
\begin{align*}
\Theta_n(l,m|2-\al)
=\Theta_n(m,l|\al).
\end{align*}
This implies the symmetry between the first and
the second chiralities: the roles of 
$\betab^*_{2j-1}$, $\betab^*_{\mathrm{screen},j}$ and $\betab^*_{\mathrm{SCREEN},j}$
are interchanged with 
$\bar{\gammab}^*_{2j-1}$, $\bar{\gammab}^*_{\mathrm{screen},j}$ and 
$\bar{\gammab}^*_{\mathrm{SCREEN},j}$, respectively,
and vice versa for $\gammab^*$ and $\bar\betab^*$.
The expansions at $\infty$ and $0$ are interchanged:
\begin{align*}
&\bar\betab^*(\la)=\sum_{j=1}^\infty\la^{\frac{2j-1}\nu}\bar\betab^*_{2j-1},\
\bar\gammab^*(\la)=\sum_{j=1}^\infty\la^{\frac{2j-1}\nu}\bar\gammab^*_{2j-1},\\
&\bar\betab^*_{\rm screen}(\la)=\sum_{j=1}^\infty\la^{-2j+\al}\bar\betab^*_{{\rm screen},j},\
\bar\gammab^*_{\rm screen}(\la)=\sum_{j=1}^\infty\la^{-2j+2-\al}\bar\gammab^*_{{\rm screen},j},\\
&\bar\betab^*_{\rm SCREEN}(\la)
=\sum_{j=1}^\infty\la^{\frac{2j-2+\al}{1-\nu}}\bar\betab^*_{{\rm SCREEN},j},\
\bar\gammab^*_{\rm SCREEN}(\la)
=\sum_{j=1}^\infty\la^{\frac{2j-\al}{1-\nu}}\bar\gammab^*_{{\rm SCREEN},j}.
\end{align*}
However, since the relations between $\la,\mu$ and $t,u$ are reversed in power,
in terms of the integration variables $l,m$ in \eqref{omega-bar}, there is no interchange
between the upper and lower half planes wherein we take residues.
The domain of convergence for the integral representation of the function
$\omega^{\rm sc}_0$ is given by \eqref{CONVDOM}.
It will change for $\bar\omega_0^{\rm sc}$ because of the change
\begin{align}
\bigl(e^{\pi i \nu+\delta}t\bigr)^{il}\rightarrow\bigl(e^{-\pi i \nu+\delta}t\bigr)^{il}
\label{CHANGESIGN}
\end{align}
in the integrand. However, by the same reason as above,
the domain of convergence in the variable $\lambda^2$ is unchanged.
There is a sign change in the residues caused by the change \eqref{CHANGESIGN}.
This will bring the pairings
\begin{align*}
&\langle\bar\betab^*_{a}\bar\gammab^*_{b}\rangle_{\kappa,\al}=
\tilde D_{a}(2-\al)\tilde D_{b}(\al)\tilde\Theta({\textstyle \frac{ib}{2\nu},\frac{ia}{2\nu}}|\kappa,\al),\\
&\langle\bar\betab^*_{{\rm screen},j}\bar\gammab^*_{b}\rangle_{\kappa,\al}=-
\tilde E'_j(\al)\tilde D_{b}(\al)\tilde\Theta({\textstyle \frac{ib}{2\nu},-i(j-\frac\al2)}|\kappa,\al),\\
&\tilde E'_k(\al)=-\frac1{\sqrt{i}}\frac{(-1)^k}{(k-1)!}\frac{\Gamma(\frac12+\nu(k-\frac\al2))}
{\Gamma(1-(1-\nu)k-\frac{\nu\al}2)}\cdot e^{-\pi i\nu(k-\frac\al2)}.
\end{align*}
The modification of the kernel function in the integral
for the second screening operator in the anti-chiral case is $\tilde S(l,\al)\rightarrow\tilde S(l,\al)\bar A_-(l,\al)$
where
\begin{align*}
\bar A_-(l,\al)=-\frac{\sinh(\pi \nu l)}{\sinh\pi\bigl((1-\nu)l+i\frac{\al}{2}\bigr)}e^{-\pi(l+\frac {i\al} 2)}\,.
\end{align*}
This will bring
\begin{align*}
&\bar\betab^*_-(\la)=\tilde{\bar\betab}^*(\la)+\bar\betab^*_{\rm SCREEN}(\al),\
\bar\gammab^*_-(\la)=\tilde{\bar\gammab}^*(\la)+\bar\gammab^*_{\rm SCREEN}(\al),\\
&\tilde{\bar\betab}^*(\la)=\sum_{j=1}^\infty \la^{\frac{2j-1}\nu}
\frac{e^{-\frac{\pi i}{2\nu}(2j-1+\nu\al)}}{\cos\frac{\pi}{2\nu}(2j-1+\nu\al)}
\cdot\bar\betab^*_{2j-1},\\
&\tilde{\bar\gammab}^*(\la)=\sum_{j=1}^\infty \la^{\frac{2j-1}\nu}
\frac{e^{-\frac{\pi i}{2\nu}(2j-1-\nu\al)}}{\cos\frac{\pi}{2\nu}(2j-1-\nu\al)}
\cdot\bar\gammab^*_{2j-1},\\
&\langle\bar\betab^*_{a}\bar\gammab^*_{{\rm SCREEN},k}\rangle_{\kappa,\al}=-
\tilde D_{a}(2-\al)\tilde F_k(\al)\tilde\Theta({\textstyle \frac i{1-\nu}(k-\frac\al2),\frac{ia}{2\nu}}|\kappa,\al),\\
&\langle\bar\betab^*_{{\rm screen},j}\bar\gammab^*_{{\rm SCREEN},k}\rangle_{\kappa,\al}=
\tilde E'_j(\al)\tilde F_k(\al)\tilde\Theta({\textstyle \frac i{1-\nu}(k-\frac\al2),-i(j-\frac\al2)}|\kappa,\al).
\end{align*}
\subsection{Fermionic descendants}\label{sec:CFTdes}
Let us formulate our conjectures 
relating various 
primary fields with shifted exponents and  their fermionic descendants. 
We follow the multi-index notation in \cite{HGSV}
and write, for 
$I=\{i_1,\cdots,i_p\}$ ($i_1<\cdots<i_p$), 
\begin{align*}
&\betab^*_{I}=\betab^*_{i_1}\cdots \betab^*_{i_p}\,,
\quad
\gammab^*_{I}=\gammab^*_{i_p}\cdots \gammab^*_{i_1}\,,
\end{align*}
arranging $\betab^*$'s from left to right and $\gammab^*$'s in the opposite order. 
We also set 
\begin{align*}
&I(n):=\{1,2,\cdots,n\}\,,
\quad I_\mathrm{odd}(n):=\{1,3,\cdots,2n-1\}\,,
\end{align*}
and write $aI+b$ to mean the set $\{ai+b \mid i\in I\}$.

For the first screening operators the conjectures in \cite{HGSV} state as follows. 
Let $n$ be an non-negative integer,  and let 
$I^\pm\subset 2\Z_{\ge0}+1$ be multi-indices such that $\#(I^+)=\#(I^-)$.
Then 
\begin{align}
&\phi_{\al+2n\frac{1-\nu}{\nu}}(0)
\simeq c_{n,0}(\al)
\betab^*_{I_{\mathrm{odd}}(n)}\gammab^*_{\mathrm{screen},I(n)}
\phi_\al(0)\,,
\label{chiral1}
\\
&\betab^*_{I^+}\gammab^*_{I^-}\phi_{\al+2n\frac{1-\nu}{\nu}}(0)
\simeq c_{n,0}(\al)
\betab^*_{I^++2n}\gammab^*_{I^--2n}
\betab^*_{I_{\mathrm{odd}}(n)}\gammab^*_{\mathrm{screen},I(n)}\phi_\al(0)\,.
\label{chiral2}
\end{align}
Here the sign $\simeq$ means both sides have the same three point functions, e.g., 
\begin{align*}
&\langle 1-\kappa | \betab^*_{I^+}\gammab^*_{I^-}
\phi_{\al+2n\frac{1-\nu}{\nu}}(0)|1+\kappa \rangle
\\
&\quad =c_{n,0}(\al)
\langle 1-\kappa | \betab^*_{I^++2n}\gammab^*_{I^--2n}
\betab^*_{I_{\mathrm{odd}}(n)}\gammab^*_{\mathrm{screen},I(n)}\phi_\al(0)
|1+\kappa \rangle\,.
\end{align*}
The main requirement is that in the two sides the $\kappa$ dependent factors must coincide. 
The proportionality constants $c_{n,0}(\al)$ 
depend on the choice of normalisation of the primary fields.  
We shall fix them in the next subsection. 

In the right hand side of \eqref{chiral2}, fermions with negative indices may appear. 
They are defined to be annihilation operators by enforcing the rule
\begin{align}
&\betab^*_{-a}=t_a(2-\al)\gammab_{a}\,,\quad \gammab^*_{-a}=-t_a(\al)\betab_{a}\,,\label{annih}\\
&[\betab_a,\betab^*_b]_+=\delta_{a,b},\quad[\gammab_a,\gammab^*_b]_+=\delta_{a,b}\,,\nn
\end{align}
where $t_a(\al)$ is defined in \eqref{tal}.

Under the exchange $1-\nu\leftrightarrow 1/(1-\nu)$, 
$\betab^*_{\mathrm{screen}}$ and $\gammab^*_{\mathrm{screen}}$
interchange their roles with
$\gammab^*_{\mathrm{SCREEN}}$ and $\betab^*_{\mathrm{SCREEN}}$, respectively.  
Hence one expects the following relations:
\begin{align}
&\phi_{\al-2m\frac{1}{\nu}}(0)
\simeq c_{0,m}(\al)
\gammab^*_{I_{\mathrm{odd}}(m)}\betab^*_{\mathrm{SCREEN},I(m)}\phi_\al(0)\,,
\label{chiral3}
\\
&\betab^*_{I^+}\gammab^*_{I^-}\phi_{\al-2m\frac{1}{\nu}}(0)
\simeq c_{0,m}(\al)\betab^*_{I^+-2m}\gammab^*_{I^-+2m}
\gammab^*_{I_{\mathrm{odd}}(m)}\betab^*_{\mathrm{SCREEN},I(m)}\phi_\al(0)\,.
\label{chiral4}
\end{align}
Obviously the scaling dimensions match in both sides.

More generally, set
\begin{align}
&\al(n,m)=\al+2n{\textstyle\frac{1-\nu}\nu}-2m{\textstyle\frac1\nu}\,.
\label{alnm}
\end{align}
Combining \eqref{chiral1}--\eqref{chiral4} together and using \eqref{annih},   
one finds that 
\begin{align}
\betab^*_{I^+}\gammab^*_{I^-}\phi_{\al(n,m)}(0)
&\simeq c_{n,m}(\al)
\begin{cases}
\betab^*_{I^++2(n-m)}\gammab^*_{I^--2(n-m)}\betab^*_{I_\mathrm{odd}(n-m)}\phi^{(n,m)}_\al(0)
&(n>m)\,,
\\
\betab^*_{I^+}\gammab^*_{I^-}\phi^{(n,m)}_\al(0)&(n=m)\,,\\
\betab^*_{I^+-2(m-n)}\gammab^*_{I^-+2(m-n)}
\gammab^*_{I_\mathrm{odd}(m-n)}
\phi_\al^{(n,m)}(0)&(n<m)\,,\\
\end{cases}
\label{chiral5}
\end{align}
where we have set 
\begin{align*}
\phi_\al^{(n,m)}(0)=
\betab^*_{\mathrm{SCREEN},I(m)}\gammab^*_{\mathrm{screen},I(n)}\phi_\al(0)\,.
\end{align*}
Note that, in the case $m=n$, the $\betab^*_{2j-1}$, $\gammab^*_{2j-1}$ 
drop out altogether from the primary field, leaving 
\begin{align*}
\phi_{\al(m,m)}(0)\simeq c_{m,m}(\al)\phi^{(m,m)}_\al(0)\,.
\end{align*}


\subsection{Consistency check for descendants}
Formula \eqref{chiral5} requires some consistency conditions, which determine
$c_{n,m}(\al)$ once we fix $c_{1,0}(\al)$ and $c_{0,1}(\al)$. We show (modulo computer checks)
that if we fix $c_{1,0}(\al)$, $c_{0,1}(\al)$ and $c_{1,1}(\al)$, then, $c_{n,m}(\al)$ is given by
\begin{align}
&c_{n,m}(\al)=
\begin{cases}
(-1)^{m(n-m)}
\prod_{j=0}^{n-m-1}c_{1,0}(\al(m+j,m))\prod_{j=0}^{m-1}c_{1,1}(\al(j,j))
&\text{ if $n\geq m$};\label{NORMALIZATION}\\
(-1)^{n(m-n))}
\prod_{j=0}^{m-n-1}c_{0,1}(\al(n,n+j))\prod_{j=0}^{n-1}c_{1,1}(\al(j,j))
&\text{ if $n\leq m$}.\\
\end{cases}
\end{align}

Recall the reduction rule
\begin{align*}
&\mathcal F(\betab^*_{a},\gammab^*_{b},
\betab^*_{{\rm SCREEN},j},\gammab^*_{{\rm screen},k})\phi_{\al(n,m)}\\
&\quad\simeq\mathcal F(\betab^*_{a+2(n-m)},\gammab^*_{b+2(m-n)},
\betab^*_{{\rm SCREEN},j+m},\gammab^*_{{\rm screen},k+n})\\
&\qquad\times
\begin{cases}
\betab^*_{I_{\rm odd}}(n-m)&(n\geq m)\\
\gammab^*_{I_{\rm odd}}(m-n)&(n\leq m)
\end{cases}
\Bigg\}\cdot
\betab^*_{{\rm SCREEN},I(m)},\gammab^*_{{\rm screen},I(n)})
\phi_\al,
\end{align*}
where $\mathcal F$ is a monomial of the fermions.
There are consistency conditions among the three quantities
$c_{1,0}(\al)$, $c_{0,1}(\al)$ and $c_{1,1}(\al)$.
One can reduce $\phi_{\al(1,1)}$ to $\phi_\al$ in three different ways.
First we have
\begin{align*}
\phi_{\al(1,1)}\simeq c_{1,1}(\al)
\betab^*_{{\rm SCREEN},1}\gammab^*_{{\rm screen},1}\phi_\al
\end{align*}
We have also
\begin{align*}
\phi_{\al(1,1)}&=c_{1,0}(\al(0,1))\betab^*_1\gammab^*_{{\rm screen},1}\phi_{\al(0,1)}\\
&=c_{1,0}(\al(0,1))c_{0,1}(\al)
\betab^*_{-1}\gammab^*_{{\rm screen},1}\gammab^*_1\betab^*_{{\rm SCREEN},1}\phi_\al\\
&=\frac i\nu\cot\frac\pi{2\nu}(1-\nu\al)c_{1,0}(\al(0,1))c_{0,1}(\al)
\betab^*_{{\rm SCREEN},1}\gammab^*_{{\rm screen},1}\phi_\al
\end{align*}
This and a similar calculation in a reversed order give the compatibility condition
\begin{align*}
c_{1,1}(\al)&=\frac i\nu\cot\frac\pi{2\nu}(1-\nu\al)c_{1,0}(\al(0,1))c_{0,1}(\al)\\
&=\frac i\nu\cot\frac\pi{2\nu}(1+\nu\al)c_{0,1}(\al(1,0))c_{1,0}(\al).
\end{align*}
A solution to the compatibility condition is given by
\begin{align*}
c_{1,0}(\al)&=1/ \cos\pi ({\textstyle\frac\al2+\frac1{2\nu}}),\\
c_{0,1}(\al)&=\sin\pi ({-\textstyle\frac\al2+\frac1{2\nu}}).
\end{align*}
Now let us discuss $c_{n,m}(\al)$ to be given by \eqref{NORMALIZATION}.
Let us choose $I^+,I^-\subset2\Z_{\geq0}+1$ such that 
$$\#(I^+)=\#(I^-)=L,$$
and consider the descendant $\beb^*_{I^+}\gab^*_{I^-}\phi_{\al(n,m)}$.
Let us use the abbreviation $$\langle\ast\rangle_\kappa=\langle 1-\kappa|\ast|1+\kappa\rangle.$$
One can compute the ratio of 3 point functions
$\langle \betab^*_{I^+}\gammab^*_{I^-}\phi_{\al(n,m)}\rangle_\kappa/
\langle \phi_\al\rangle_\kappa$ in two different ways.  
For instance, if $n\geq m$, then \eqref{chiral5} gives the consistency condition
\begin{align*}
&\frac{\langle \betab^*_{I^+}\gammab^*_{I^-}\phi_{\al(n,m)}\rangle_\kappa}
{\langle \phi_{\al(n,m)}\rangle_\kappa}
\prod_{j=0}^{n-m-1}
\frac{\langle \phi_{\al(m+j+1,m)}\rangle_\kappa}
{\langle \phi_{\al(m+j,m)}\rangle_\kappa}
\prod_{j=0}^{m-1}
\frac{\langle \phi_{\al(j+1,j+1)}\rangle_\kappa}
{\langle \phi_{\al(j,j)}\rangle_\kappa}
\\
&=c_{n,m}(\al)
\frac{\langle \betab^*_{I^++2(n-m)}\gammab^*_{I^--2(n-m)}\betab^*_{I_\mathrm{odd}(n-m)}
\gammab^*_{\mathrm{SCREEN},I(m)}\gammab^*_{\mathrm{screen},I(n)}
\phi_{\al}\rangle_\kappa}
{\langle \phi_{\al}\rangle_\kappa}\,,
\end{align*}
and similarly for $m\le n$. One can apply \eqref{chiral5} to each term in the product of the left hand side.
Then, calculation of the two sides by Wick contraction
leads to a family of identities involving the functions $\tilde\Theta(l,m|\kappa,\al)$ and
$\tilde D_j(\al)$, $\tilde E_j(\al)$, $\tilde F_j(\al)$.

Two remarks are in order. First,
it is easy to see that the consistency condition is unchanged if we replace $\tilde\Theta(l,m|\kappa,\al)$
with $\Theta(l,m|\kappa,\al)$. This is because $x(\kappa)$ appears with the level as exponent.
So, we forget $x(\kappa)$ in the following calculation.
Next, we have performed computer checks with various choices of $n,m$
that the consistency condition reduces to that of $\kappa=\infty$. We did such checks
up to order $\kappa^{-6}$. Note that
\begin{align*}
\Theta(l,m|\infty,\al)=-\frac i{l+m}.
\end{align*}
In the following we use only this specialization, and 
write
$\Theta(l,m|\infty,\al)$ simply as $\Theta(l,m)$.

In order to write out these relations systematically, 
we introduce the following index sets:
\begin{align*}
I^\pm_>(n,m)&=\{j\pm2(n-m)|j\in I^\pm\}\cap\Z_{>0},\\
I^\pm_c(n,m)&=\{-j\mp2(n-m)|j\in I^\pm\}\cap\Z_{>0},\\
\tilde I^\pm(n,m)&=\left(I_{\rm odd}\bigl(\pm(n-m)\bigr)
\backslash I^\mp_c(n,m)\right)\sqcup I^\pm_>(n,m),
\end{align*}
where $I_{\rm odd}(n)$ is the empty set if $n\leq0$.
The subscript $c$ in $I^\pm_c(n,m)$ signifies the fermion contraction.
Note that these index sets are actually depend only on the difference $n-m$.

For example, $I^\pm_c(0,0)=\emptyset,\tilde I^\pm(0,0)=I^\pm$;
if $I^+=\{1,3\}$ and $I^-=\{1,5\}$ we have
$\tilde I^+(2,0)=\{1,5,7\}$, $\tilde I^-(2,0)=\{1\}$, $I^+_c(2,0)=\emptyset$, $I^-_c(2,0)=\{3\}$.
Note that if $n\geq m$, then $I^+_c(n,m)=\emptyset$; if $n\leq m$, then $I^-_c(n,m)=\emptyset$.

Set
 \begin{align*}
&\vartheta_{n,m}=\sum_{2k-1\in I^+_c(n,m)\sqcup I^-_c(n,m)}(k-1),\\
&t^\pm_{n,m}(\al)=\prod_{a\in I^\pm_c(n,m)}{\textstyle\frac i\nu}\cot{\textstyle\frac\pi{2\nu}(a\mp\nu\al)}.
\end{align*}
We define the Cauchy determinant with normalization factors
 \begin{align*}
&{\mathcal D}_{I^+,I^-}(n,m|\al)
\\
&=\det 
\begin{pmatrix}
\left(\Theta(\frac{ia}{2\nu},\frac{ib}{2\nu})\right)_{a\in \tilde I^+(n,m),b\in\tilde I^-(n,m)}&
\left(\Theta(\frac{ia}{2\nu},-\frac{i(2k-\al)}{2})\right)_{a\in \tilde I^+(n,m),k\in I(n)}\\[5pt]
\left(\Theta(\frac{i(2j-\al)}{2(1-\nu)},\frac{ib}{2\nu})\right)_{j\in I(m),b\in\tilde I^-(n,m)}&
\left(\Theta(\frac{i(2j-\al)}{2(1-\nu)},-\frac{i(2k-\al)}{2})\right)_{j\in I(m),k\in I(n)}
\end{pmatrix}\\
&\times\prod_{a\in\tilde I^+(n,m)}\tilde D_{a}(\al)
\prod_{a\in\tilde I^-(n,m)}\tilde D_{a}(2-\al)
\prod_{j\in I(m)}\tilde F_j(\al)\prod_{k\in I(n)}\tilde E_k(\al).
\end{align*}
In this notation, we have
\begin{align*}
\frac{\langle\betab^*_{I^+}\gammab^*_{I^-}\phi_{\al(n,m)}(0)\rangle_{\kappa=\infty}}
{\langle\phi_\al(0)\rangle_{\kappa=\infty}}
&=c_{n,m}(\al)
\begin{cases}
(-1)^{L(n-m)+\vartheta_{n,m}}t^-_{n,m}(\al){\mathcal D}_{I^+,I^-}(n,m|\al)
&(n\geq m);\\
(-1)^{(L+1)(m-n)+\vartheta_{n,m}}t^+_{n,m}(\al)
{\mathcal D}_{I^+,I^-}(n,m|\al)&(n\leq m).
\end{cases}
\end{align*}
On the other hand, the left hand side of the consistency relation reads
\begin{align*}
&\frac{\langle\betab^*_{I^+}\gammab^*_{I^-}\phi_{\al(n,m)}(0)\rangle_\infty}
{\langle\phi_{\al(n,m)}(0)\rangle_\infty}
={\mathcal D}_{I^+,I^-}(0,0|\al(n,m)),
\end{align*}
\begin{align*}
&\frac{\langle\phi_{\al(n,m)}(0)\rangle_\infty}{\langle\phi_\al(0)\rangle_\infty}
=\prod_{j=1}^N\Theta\Bigl(i{\textstyle\frac{j-\frac\al 2}{1-\nu},-i\left(j-\frac\al 2\right)}\Bigr)
\prod_{j=1}^{m-N}
\Theta\Bigl({\textstyle\frac i{1-\nu}\bigl(1-\frac{\al(N,N+j)}2\bigr)},i{\textstyle\frac1{2\nu}}\Bigr)
\end{align*}
\begin{align*}
&\times\prod_{j=1}^{n-N}
\Theta\Bigl(i{\textstyle\frac1{2\nu}},-i{\textstyle\bigl(1-\frac{\al(N+j,N)}2\bigr)}\Bigr)
\prod_{j=0}^{N-1}\tilde E_1(\al(j,j))\tilde F_1(\al(j,j))
\end{align*}
\begin{align*}
&\times\prod_{j=0}^{n-N-1}\tilde D_1(\al(N+j,N)\tilde E_1(\al(N+j,N))\\
&\times\prod_{j=0}^{m-N-1}(-1)\tilde D_1(2-\al(N,N+j))
\tilde F_1(\al(N,N+j))\\
&\times
\begin{cases}
\prod_{j=0}^{n-m-1}c_{1,0}(\al(m+j,m))\prod_{j=0}^{m-1}c_{1,1}(\al(j,j))
&\text{ if $n\geq m$};\\
\prod_{j=0}^{m-n-1}c_{0,1}(\al(n,n+j))\prod_{j=0}^{n-1}c_{1,1}(\al(j,j))
&\text{ if $n\leq m$},
\end{cases}
\end{align*}
where $N=\min(n,m)$.
The formula \eqref{NORMALIZATION} for the normalization factor $c_{n,m}(\al)$
is obtained from these relations.

The relations for the normalization factor \eqref{NORMALIZATION} are the same for
the anti-chiral case. This follows from the formulas for the fermionic pairings that are given in Section 2.2.

\subsection{Gluing two chiralities}
The primary field in the full CFT is 
\begin{align*}
\Phi_\al(0)=S(\al)\phi_\al(0)\bar{\phi}_{\al}(0)
\end{align*}
with some numerical coefficient $S(\al)$. 

Introduce the screened primary field by 
\begin{align}
\Phi_\al^{(n,m)}(0)&=s_{n,m}(\al)
\betab^*_{\mathrm{SCREEN},I(m)}\bar{\betab}^*_{\mathrm{screen},I(n)}
\gammab^*_{\mathrm{screen},I(n)}\bar{\gammab}^*_{\mathrm{SCREEN},I(m)}\Phi _\al(0)\,,
\label{ScPri}\\
s_{n,m}(\al)&=\mub^{2n-\frac 2 {1-\nu}m}
\Bigl\{\prod_{k=1}^nt'_k(\al)\prod_{j=1}^m t''_j(\al)
\Bigr\}^{-1}\,.
\nn
\end{align}
Gluing toghether the relation \eqref{chiral5} with their anti-chiral partners, 
one finds that 
\begin{align}
\Phi_{\al(n,m)}(0)
\cong 
C_{n,m}(\al)\times
\begin{cases}
\betab ^*_{I_\mathrm{odd}(n-m)}\bar{\gammab} ^*_{I_\mathrm{odd}(n-m)}\Phi _\al^{(n,m)}(0)
&(n>m)\,,\\
\Phi _\al^{(m,m)}(0) &(n=m)\,, \\ 
\bar{\betab} ^*_{I_\mathrm{odd}(m-n)}
\gammab ^*_{I_\mathrm{odd}(m-n)}\Phi _\al^{(n,m)}(0)
&(n<m)\,,\\
\end{cases}
\label{full}
\end{align}
where the normalisation coefficient $C_{n,m}(\al)$ is related to $c_{n,m}(\al)$ in 
\eqref{chiral5} and its anti-chiral counterpart $\bar{c}_{n,m}(\al)$ via
\begin{align*}
C_{n,m}(\al)s_{n,m}(\al)
=\frac{S(\al(n,m))}{S(\al)}c_{n,m}(\al)\bar{c}_{n,m}(\al)\,.
\end{align*}
The recursive relations, which determine the general case
out of $C_{0,1},C_{1,0},C_{1,1}$, are 
\begin{align*}
&C_{n,m}(\al)=
\begin{cases}
C_{n,m-1}(\al)C_{0,1}\Bigl(\al(n,m-1)\Bigr)&(n<m)\,,\\
C_{m-1,m-1}(\al)C_{1,1}\Bigl(\al(m-1,m-1)\Bigr)& (n=m)\,,\\
C_{n-1,m}(\al)C_{1,0}\Bigl(\al(n-1,m)\Bigr)&(n>m)\,.\\
\end{cases}
\end{align*}
These relations are deduced from the corresponding ones for
$c_{n,m}(\al),\bar{c}_{n,m}(\al)$ and the simple relations
\begin{align*}
t'_n(\al)=t'_1(\al(n-1,m)),\
t''_m(\al)=t''_1(\al(n,m-1)).
\end{align*}

Choosing the `exponential normalisation' 
$\Phi_\al(z,\bar{z})\Phi_{-\al}(0)=|z|^{-4\Delta_{\al}}
{(1+O(|z|^2))}$, 
one can determine $C_{n,m}(\al)$ by comparing with the Liouville 
three point functions by Zamolodchikov-Zamolodchikov \cite{ZZ3point}. 

This calculation for the 
coefficient $C_{1,0}(\al)$ is explained in \cite{HGSV}, eq.(6.8). 
For $C_{0,1}(\al)$, the relevant formula can be obtained by the substitution 
$b\to b^{-1}$ in \cite{HGSV}, (A.2) (which amounts to 
$-1/\nu\to (1-\nu)/\nu$
), and comparing the result 
with the pairing $\langle \betab^*_{\mathrm{SCREEN},1}\gammab^*_1\rangle_{\kappa,\al}$
given by \eqref{FPAIRING}.

We have checked that the $\kappa$-dependent part agrees up to the degree $\kappa^{-6}$. 
Collecting the coefficients, we obtain the following result. 
\begin{align}
&C_{1,0}(\al)=\Gamma(\nu)^{2\al+2\frac{1-\nu}{\nu}}\,
i\nu \tan\frac{\pi}{2\nu}(\nu\al+1)
\label{C01}\\
&\quad\times 
\frac{\Gamma\Bigl(\frac{1}{2}-\frac{\al}{2}-\frac{1-\nu}{2\nu}\Bigr)}
{\Gamma\Bigl(\frac{1}{2}+\frac{\al}{2}+\frac{1-\nu}{2\nu}\Bigr)}
\frac{\Gamma\Bigl(\frac{\al}{2}+\frac{1-\nu}{2\nu}\Bigr)}
{\Gamma\Bigl(-\frac{\al}{2}-\frac{1-\nu}{2\nu}\Bigr)}
\frac{\Gamma\Bigl(-1+\nu-\nu\al\Bigr)}
{\Gamma\Bigl(1-\nu+\nu\al\Bigr)}\,,
\nn\\
&C_{0,1}(\al)=-\{(1-\nu)^{2\nu}\Gamma(\nu)^{-2}\}^{\frac{\nu\al-1}{\nu(1-\nu)}}\,
i\nu\tan\frac{\pi}{2\nu}(\nu\al-1)
\label{C10}\\
&\quad\times
\frac{\Gamma\Bigl(\frac{1}{2}-\frac{\al}{2}+\frac{1}{2\nu}\Bigr)}
{\Gamma\Bigl(\frac{1}{2}+\frac{\al}{2}-\frac{1}{2\nu}\Bigr)}
\frac{\Gamma\Bigl(\frac{\al}{2}-\frac{1}{2\nu}\Bigr)}
{\Gamma\Bigl(-\frac{\al}{2}+\frac{1}{2\nu}\Bigr)}
\frac{\Gamma\Bigl(-\frac{1}{1-\nu}+\frac{\nu\al}{1-\nu}\Bigr)}
{\Gamma\Bigl(\frac{1}{1-\nu}-\frac{\nu\al}{1-\nu}\Bigr)}\,,
\nn
\end{align}
and
\begin{align}
&C_{1,1}(\al)=\{(1-\nu)^2 \Gamma(\nu)^{-2}\}^{\frac{\nu(\al-1)}{1-\nu}}
\times
\frac{\Gamma\Bigl(\frac{\nu}{1-\nu}(\al-1)\Bigr)}
{\Gamma\Bigl(-\frac{\nu}{1-\nu}(\al-1)\Bigr)}
\frac{\Gamma\bigl(-\nu(\al-1)\bigr)}{\Gamma\bigl(\nu(\al-1)\bigr)}\,.
\label{C11}
\end{align}
The third one can be obtained from the first two by the relation
\begin{align*}
C_{1,1}(\al)=-{\textstyle\frac1{\nu^2}}\cot^2{\textstyle\frac\pi2(\al-\frac1\nu)}
C_{0,1}(\al)C_{1,0}({\textstyle\al-\frac2\nu}).
\end{align*}

\section{Sine-Gordon model}
In this section, we discuss the sine-Gordon model. 
Since the working is entirely similar to the one of \cite{HGSV}, 
we shall be rather brief, indicating only the necessary formulas. 

\subsection{Fermions in sine-Gordon model}
As in the CFT case, the fermions in the sine-Gordon model are defined
by giving the expectation values of basic descendants.
The way of introducing the second screening operators is similar to that in CFT: 
in place of the function 
$\omega^\mathrm{sG}_{R}(\zeta,\xi|\al)$
given in \cite{HGSV}, eq. (7.8), 
we use the following formal integral
\begin{align*}
&\omega^\mathrm{sG}_{R,\epsilon,\epsilon'}(\zeta,\xi|\al)
=-\frac{\pi i}{2}\int\!\!\int\frac{dl}{2\pi}\frac{dm}{2\pi}\,
\zeta^{2il}\xi^{2im}\,\Theta^\mathrm{sG}_{R}(l,m|\al)
\\
&\quad\times
A_\epsilon(l,\al)\frac{e^{-\pi\nu l}}{\cosh\pi\nu l}\,
A_{\epsilon'}(m,2-\al)\frac{e^{-\pi\nu m}}{\cosh\pi\nu m}\,,
\end{align*}
which we understand as the expansions in $\zeta^2,\xi^2$ at $\infty,0$;
the expansions are obtained by taking residues in $l,m$ in the upper/lower half planes.
Accordingly, we define $\betab^{+*}_\epsilon(\zeta),\gammab^{+*}_\epsilon(\zeta)\ (\epsilon=\pm)$
for $\zeta^2\rightarrow\infty$, and
$\betab^{-*}_\epsilon(\zeta),\gammab^{-*}_\epsilon(\zeta)\ (\epsilon=\pm)$
for $\zeta^2\rightarrow0$ as follows:
\begin{align*}
\betab^{+*}_\epsilon(\zeta)=
\begin{cases}
\betab^*(\mub \zeta)+\bar{\betab}^*_{\mathrm{screen}}(\mub^{-1}\zeta)& (\epsilon=+);\\
\tilde{\betab}^*(\mub \zeta)+\betab^*_{\mathrm{SCREEN}}(\mub\zeta)
+\bar{\betab}^*_{\mathrm{screen}}(\mub^{-1}\zeta)& (\epsilon=-),\\
\end{cases}\\
\gammab^{+*}_\epsilon(\zeta)=
\begin{cases}
\gammab^*(\mub \zeta)+\bar{\gammab}^*_{\mathrm{screen}}(\mub^{-1}\zeta)
&  (\epsilon=+);\\\tilde{\gammab}^*(\mub \zeta)+\gammab^*_{\mathrm{SCREEN}}(\mub\zeta)
+\bar{\gammab}^*_{\mathrm{screen}}(\mub^{-1}\zeta)&(\epsilon=-),\\
\end{cases}\\
\betab^{-*}_\epsilon(\zeta)=
\begin{cases}
\bar{\betab}^*(\mub^{-1} \zeta)+\betab^*_{\mathrm{screen}}(\mub\zeta)
& (\epsilon=+);\\\tilde{\bar{\betab}}^*(\mub^{-1} \zeta)
+\bar{\betab}^*_{\mathrm{SCREEN}}(\mub^{-1}\zeta)
+\betab^*_{\mathrm{screen}}(\mub\zeta)& (\epsilon=-),\\
\end{cases}\\
\gammab^{-*}_\epsilon(\zeta)=
\begin{cases}
\bar{\gammab}^*(\mub^{-1} \zeta)+\gammab^*_{\mathrm{screen}}(\mub\zeta)
&  (\epsilon=+);\\\tilde{\bar{\gammab}}^*(\mub^{-1} \zeta)
+\bar{\gammab}^*_{\mathrm{SCREEN}}(\mub^{-1}\zeta)
+\gammab^*_{\mathrm{screen}}(\mub\zeta)&(\epsilon=-).\\
\end{cases}
\end{align*}
The prescription for the fermion pairings is as follows. 
\begin{align*}
&\langle \betab^{\pm*}_\epsilon(\zeta)\gammab^{\pm*}_{\epsilon'}(\xi)\rangle^{\mathrm{sG}}_R
=\omega^{\mathrm{sG}}_{R,\epsilon,\epsilon'}(\zeta,\xi|\al)\,,\\
&\langle \betab^{\pm*}_\epsilon(\zeta)\gammab^{\mp*}_{\epsilon'}(\xi)\rangle^{\mathrm{sG}}_R
=\omega^{\mathrm{sG}}_{R,\epsilon,\epsilon'}(\zeta,\xi|\al)
+\omega_{0,\epsilon,\epsilon'}(\zeta/\xi|\al).
\end{align*}
It is immediate to write them in terms of the Fourier components. 
For that purpose let us introduce the shorthand notation
\begin{align*}
&\hat{\betab}^*_a=\begin{cases}
\mub^{-\frac{a}{\nu}}\betab^*_a & (a>0),\\
-\mub^{\frac{a}{\nu}}\bar{\betab}^*_{-a}& (a<0),\\
\end{cases}
\quad
\hat{\gammab}^*_a=\begin{cases}
\mub^{-\frac{a}{\nu}}\gammab^*_a & (a>0),\\
-\mub^{\frac{a}{\nu}}\bar{\gammab}^*_{-a}& (a<0),\\
\end{cases}
\end{align*}
\begin{align*}
&\hat{\betab}^*_{\mathrm{SCREEN},j}=
\begin{cases}
\mub^{-\frac{2j-\al}{1-\nu}}\betab^*_{\mathrm{SCREEN},j} & (j>0),\\
-\mub^{\frac{2j-\al}{1-\nu}}\bar{\betab}^*_{\mathrm{SCREEN},1-j}& (j\le 0).
\end{cases}
\end{align*}
\begin{align*}
&\hat{\gammab}^*_{\mathrm{SCREEN},j}=
\begin{cases}
\mub^{-\frac{2j-2+\al}{1-\nu}}\gammab^*_{\mathrm{SCREEN},j} & (j>0),\\
-\mub^{\frac{2j-2+\al}{1-\nu}}\bar{\gammab}^*_{\mathrm{SCREEN},1-j}& (j\le 0).
\end{cases}
\end{align*}
The minus sign in front of $\mub$ comes from the orientation
of cycles for complex integration.

For all $a,b\in 2\Z+1$ and $j,k\in \Z$ we have
\begin{align*}
&\langle \hat{\betab}^*_a\hat{\gammab}^*_b \rangle^\mathrm{sG}_R
=\frac{i}{2\pi}\frac{1}{\nu^2}
\left\{
\Theta^{\mathrm{sG}}_R\Bigl(\frac{ia}{2\nu},\frac{ib}{2\nu}\Bigl|\al\Bigr)
+ \delta_{a+b,0}\mathrm{sgn}(a) 2\pi i\nu^2 t_a(\al)
\right\}\,,
\end{align*}
\begin{align*}
&\langle \hat{\betab}^*_a\hat{\gammab}^*_{\mathrm{SCREEN},k} \rangle^\mathrm{sG}_R
=
\frac{i}{2\pi}\frac{1}{\nu}
\Theta^{\mathrm{sG}}_R\Bigl(\frac{ia}{2\nu},\frac{i(2k+\al-2)}{2(1-\nu)}\Bigl|\al\Bigr)
\,t''_k(2-\al)\,,
\\
&\langle \hat{\betab}^*_{\mathrm{SCREEN},j}\hat{\gammab}^*_b \rangle^\mathrm{sG}_R
=
\frac{i}{2\pi}\frac{1}{\nu}
t''_j(\al)\,
\Theta^{\mathrm{sG}}_R\Bigl(\frac{i(2j-\al)}{2(1-\nu)},\frac{ib}{2\nu}\Bigl|\al\Bigr)\,,
\end{align*}
\begin{align*}
&\langle \hat{\betab}^*_{\mathrm{SCREEN},j}\hat{\gammab}^*_{\mathrm{SCREEN},k} \rangle^\mathrm{sG}_R
=
\frac{i}{2\pi}t''_j(\al)t''_k(2-\al)
\Bigl\{
\Theta^{\mathrm{sG}}_R\Bigl(\frac{i(2j-\al)}{2(1-\nu)},
\frac{i(2k+\al-2)}{2(1-\nu)}\Bigl|\al\Bigr)
\\
&\quad\qquad
-\delta_{j+k,1}\mathrm{sgn}\bigl(2j-1)
2\pi i \,t''_j(\al)^{-1}
\Bigr\}\,.
\end{align*}
The first screening operators decouple from the rest:
\begin{align}
&\langle \bar{\betab}^*_{\mathrm{screen},j}\gammab^*_\mathrm{screen,k} \rangle_R^\mathrm{sG}
=\delta_{j,k} \mub^{-2(2j-\al)}t'_j(\al)\,,\label{decouple1}\\
&\langle \betab^*_{\mathrm{screen},j}\bar{\gammab}^*_\mathrm{screen,k} \rangle_R^\mathrm{sG}
=\delta_{j,k} \mub^{-2(2j-2+\al)}t'_j(2-\al)\,.\label{decouple2}
\end{align}
It is not surprising that the symmetry between two kinds
of screening operators is broken: the sG model is a perturbation of CFT
by the operators which behaves differently with respect to them.

\subsection{One point functions on the plane}

In the limit $R\to\infty$, $\omega^{\mathrm{sG}}_R$ vanishes, and  
the only non-trivial pairings become
\begin{align*}
&\langle \betab^*_a\bar{\gammab}^*_{a} \rangle^\mathrm{sG}_\infty
=\mub^{\frac{2}{\nu}}\frac{i}{\nu}\cot\frac{\pi}{2\nu}(a+\nu\al)\,,
\quad 
\langle \bar{\betab}^*_a\gammab^*_{a} \rangle^\mathrm{sG}_\infty
=\mub^{\frac{2}{\nu}}\frac{i}{\nu}\cot\frac{\pi}{2\nu}(a-\nu\al)\,,
\\
&\langle \betab^*_{\mathrm{SCREEN},j}\bar{\gammab}^*_{\mathrm{SCREEN},j} \rangle^\mathrm{sG}_\infty
=\mub^{\frac{2}{1-\nu}(2j-\al)}t''_j(\al)\,,
\\
&\langle \bar{\betab}^*_{\mathrm{screen},j}\gammab^*_\mathrm{screen,j} \rangle_\infty^\mathrm{sG}
=\mub^{-2(2j-\al)}t'_j(\al)\,.
\end{align*}
Here again we have listed only the pairings which will be used later. 

The screened primary field is defined by the same formula
\eqref{ScPri} as before. Notice that it is normalized as 
\begin{align*}
\frac{\langle \Phi^{(n,m)}_\al(0)\rangle^{\mathrm{sG}}_\infty}
{\langle \Phi_\al(0) \rangle^{\mathrm{sG}}_\infty}
=\mub^{-2n(n-\al)+\frac{2m(m-\al)}{1-\nu}}\,.
\end{align*}
In \cite{HGSV}, we have verified for $(n,m)=(1,0)$ that this formula agrees with 
the Lukyanov-Zamolodchikov formula \cite{LukZam} for the 
one-point function on the plane. 
Let us check that it works also for the second screening operators, i.e., for 
$(n,m)=(0,1)$. 

From \eqref{ScPri} and the pairings above we calculate 
\begin{align*}
\frac{\langle \Phi_{\al-\frac{2}{\nu}}(0)\rangle^\mathrm{sG}_\infty}
{\langle\Phi_\al(0)\rangle^\mathrm{sG}_\infty}
=\mub^{\frac{2}{\nu}+\frac{2(1-\al)}{1-\nu}}
C_{0,1}(\al)\frac{i}{\nu}\cot\frac{\pi}{\nu}(1-\nu\al)\,.
\end{align*}
On the other hand, a direct computation with the formula in \cite{LukZam} gives
\begin{align*}
&\frac {\langle \Phi_{\al -\frac 2\nu}(0)\rangle^{\mathrm{sG}}_{\infty}}
{\langle \Phi_{\al }(0)\rangle^{\mathrm{sG}}_{\infty}}
=\(\mub \Gamma(\nu)\)^{\frac{2(1-\nu\al)}{\nu(1-\nu)}}
H_{0,1}\Bigl(\frac{\al}{2}-\frac{1}{2\nu}\Bigr)\,,
\\
&H_{0,1}(x)=-(1-\nu)^{\frac {4\nu}{1-\nu}x}
\frac {\Gamma(x)\Gamma(\frac 1 2-x)}{\Gamma(-x)\Gamma(\frac 1 2+x)}\cdot \frac 
{\Gamma (\frac {2\nu}{1-\nu}x)}{\Gamma (-\frac {2\nu}{1-\nu}x)}\,.
\end{align*}
We find  perfect agreement with the formula above. 

\subsection{Consistency}

Finally let us touch upon the consistency conditions.

As explained in \cite{HGSV}, section 9, the following 
identities can be derived 
from the linear integral equation defining $\Theta^{\mathrm{sG}}_R(l,m|\al)$.
\begin{align*}
&\Theta^\mathrm{sG}_R\Bigl(l,m\Bigl|\al\pm\frac{2(1-\nu)}{\nu}\Bigr)
\cdot
\Bigl(\Theta^\mathrm{sG}_R\Bigl(\pm\frac{i}{2\nu},\mp\frac{i}{2\nu}\Bigl|\al\Bigr)
-2\pi\nu\cot\frac{\pi}{2\nu}(1\pm\nu\al)
\Bigr)
\\
&=\left|
\begin{matrix}
\Theta^\mathrm{sG}_R\Bigl(\pm\frac{i}{2\nu},\mp\frac{i}{2\nu}
\Bigl|\al\Bigr)-2\pi\nu\cot\frac{\pi}{2\nu}(1\pm\nu\al)
&\Theta^\mathrm{sG}_R\Bigl(\pm\frac{i}{2\nu},m\mp\frac{i}{\nu}\Bigl|\al\Bigr)\\
\Theta^\mathrm{sG}_R\Bigl(l\pm\frac{i}{\nu},\mp\frac{i}{2\nu}\Bigl|\al\Bigr)
&\Theta^\mathrm{sG}_R\Bigl(l\pm\frac{i}{\nu},m\mp\frac{i}{\nu}\Bigl|\al\Bigr)
\\
\end{matrix}
\right|\,,
\\
&\Theta^\mathrm{sG}_R\Bigl(l,m\Bigl|\al\mp2\Bigr)
\cdot
\Bigl(\Theta^\mathrm{sG}_R\Bigl(\frac{i(1-\al\pm1)}{2(1-\nu)},
-\frac{i(1-\al\pm1)}{2(1-\nu)}\Bigl|\al\Bigr)
\mp2\pi(1-\nu)\cot\frac{\pi\nu(1-\al\pm1)}{2(1-\nu)}\Bigr)
\\
&
=\left|
\begin{matrix}
\Theta^\mathrm{sG}_R\Bigl(\frac{i(1-\al\pm1)}{2(1-\nu)},-\frac{i(1-\al\pm1)}{2(1-\nu)}
\Bigl|\al\Bigr)\mp2\pi(1-\nu)\cot\frac{\pi\nu(1-\al\pm1)}{2(1-\nu)}
&\Theta^\mathrm{sG}_R\Bigl(\frac{i(1-\al\pm1)}{2(1-\nu)},m\Bigl|\al\Bigr)
\\
\Theta^\mathrm{sG}_R\Bigl(l,-\frac{i(1-\al\pm1)}{2(1-\nu)}\Bigl|\al\Bigr) 
&\Theta^\mathrm{sG}_R\Bigl(l,m\Bigl|\al\Bigr)\\
\end{matrix}
\right|\,.
\end{align*}
Combining them together and manipulating with determinants, 
one deduces further that for all $r,s\ge 0$
\begin{align}
&\Theta^\mathrm{sG}_R\Bigl(l,m\Bigl|\al-\frac{2r}{\nu}+2s\frac{1-\nu}{\nu}\Bigr)
=\det\begin{pmatrix}
\mathcal{A} & \mathcal{B} &  \mathbf{x} \\
\mathcal{C} & \mathcal{D} & \mathbf{y} \\
{}^t\mathbf{u} & {}^t\mathbf{v}& z  \\
\end{pmatrix}
/\det\begin{pmatrix}
\mathcal{A} & \mathcal{B} \\
 \mathcal{C} & \mathcal{D} \\
\end{pmatrix}
\,,
\label{ident}
\end{align}
where
\begin{align*}
&\mathcal{A}_{j,k}=\Theta^\mathrm{sG}_R\Bigl(\frac{i(2j-\al)}{2(1-\nu)}, 
-\frac{i(2k-\al)}{2(1-\nu)}\Bigl|\al\Bigr)
-\delta_{j,k}2\pi(1-\nu)\cot\frac{\pi\nu(2j-\al)}{2(1-\nu)}\,,
\\
&\mathcal{B}_{j,b}=\Theta^\mathrm{sG}_R\Bigl(\frac{i(2j-\al)}{2(1-\nu)}, 
-\frac{ib}{2\nu}\Bigl|\al\Bigr)\,,
\quad 
\mathcal{C}_{a,k}=\Theta^\mathrm{sG}_R\Bigl(\frac{ia}{2\nu},-\frac{i(2k-\al)}{2(1-\nu)}\Bigl|\al\Bigr)\,,
\\
&\mathcal{D}_{a,b}=
\Theta^\mathrm{sG}_R\Bigl(\frac{ia}{2\nu}, -\frac{ib}{2\nu}\Bigl|\al\Bigr)
-\delta_{a,b}\mathrm{sgn}(a)\,2\pi\nu \cot\frac{\pi(a+\nu\al)}{2\nu}\,,
\\
&
\mathbf{x}_j=\Theta^\mathrm{sG}_R\Bigl(\frac{i(2j-\al)}{2(1-\nu)},m+i\frac{r-s}{\nu}\Bigl|\al\Bigr)\,,
\quad 
\mathbf{y}_a=\Theta^\mathrm{sG}_R\Bigl(\frac{ia}{2\nu},m+i\frac{r-s}{\nu}\Bigl|\al\Bigr)\,,
\\
&
\mathbf{u}_k=\Theta^\mathrm{sG}_R\Bigl(l-i\frac{r-s}{\nu}, -\frac{i(2k-\al)}{2(1-\nu)}\Bigl|\al\Bigr)\,,
\quad 
\mathbf{v}_b=\Theta^\mathrm{sG}_R\Bigl(l-i\frac{r-s}{\nu}, -\frac{ib}{2\nu}\Bigl|\al\Bigr)\,,
\\
&z=\Theta^\mathrm{sG}_R\Bigl(l-i\frac{r-s}{\nu},m+i\frac{r-s}{\nu}\Bigl|\al\Bigr)\,,
\end{align*}
and the indices range over 
\begin{align*}
j,k\in I(r), 
\quad
a,b\in 
\begin{cases}
I_\mathrm{odd}(s-r) & (s\ge r),\\
-I_{\mathrm{odd}}(r-s) & (r\ge s). \\
\end{cases}
\end{align*}

Appropriate specialisations of \eqref{ident}
guarantee the consistency of the formula \eqref{full}. 
Since the argument is the same as in \cite{HGSV}, we omit the details. 

\bigskip



\bigskip

{\it Acknowledgements.}\quad
\medskip

Research of MJ is supported by the Grant-in-Aid for Scientific 
Research B-20340027. 
Research of TM is supported by the Grant-in-Aid for Scientific Research
B-22340031.
Research of FS is supported by   SFI
under Walton Professorship scheme, by
RFBR-CNRS grant 09-02-93106
and DIADEMS program (ANR) contract number BLAN012004.
MJ and TM 
would like to thank for the hospitality extended by the Hamilton Mathematical Institute 
where a part of this work was begun. 


\end{document}